\documentclass[aps,prl,twocolumn]{revtex4}
\usepackage{bm} 
\usepackage{graphicx}
\usepackage{amsmath}
\usepackage{slashed}
\def\bea{\begin{eqnarray}}
\def\eea{\end{eqnarray}}
\def\bal{\begin{align}}
\def\eal{\end{align}}
\def\sfrac#1#2{{\textstyle \frac{#1}{#2}}}

\begin{document}
\title{\hfill {\small JLAB-THY-07-632}\\ High-precision covariant one-boson-exchange potentials\\for $np$ scattering below 350 MeV}
\author{Franz Gross}
\affiliation{Thomas Jefferson National Accelerator Facility, Newport News, VA 23606\\ College of William and Mary, Williamsburg, VA 23187}
\author{Alfred Stadler}
\affiliation{Centro de F\'isica Nuclear da Universidade de Lisboa, 1649-003 Lisboa, Portugal \\and Departamento de F\'isica da Universidade de \'Evora, 7000-671 \'Evora, Portugal}
\begin{abstract}
All realistic potential models for the two-nucleon interaction are to some extent based on boson exchange. However, in order to achieve an essentially perfect fit to the scattering data, characterized by a $\chi^2/N_\mathrm{data}\sim 1$, previous potentials have abandoned a pure one boson-exchange mechanism (OBE).  
Using a covariant theory, we have found a  OBE potential that fits the  2006 world $np$  data below 350 MeV with a $\chi^2/N_\mathrm{data} = 1.06$ for  3788 data.  Our potential has fewer adjustable parameters than previous high-precision potentials, and also reproduces the experimental triton binding energy without introducing additional irreducible three-nucleon forces.

\end{abstract}

\maketitle

A good understanding of the interaction between two nucleons is essential for the study of nuclear structure and nuclear reactions. In the long history of theoretical models of the $NN$ interaction, One-Boson-Exchange (OBE) models played a role of special importance. Yukawa's \cite{Yuk35} insight that a short-range force can be generated through the exchange of particles of finite mass led to the discovery of the pion, and later it was found that the exchange of a pion can quantitatively describe the longer-range part of the $NN$ interaction. Since the range of the force is inversely proportional to the exchanged mass, the exchange of  heavier mass bosons  generates $NN$ forces of intermediate to short range. It was found that the vector bosons $\omega$ and $\rho$ contribute to the observed spin-orbit force and strong repulsion at short internucleon distances \cite{Bri60}, and that scalar bosons provide intermediate attraction.  Today, with the development of potentials based on chiral perturbation theory (ChPT) \cite{Weinberg:1990rz}, we understand that these scalar bosons are an approximate representation of the two-pion exchange mechanism \cite{Rentmeester:2003mf}, which gives strong attraction even if there were no two-pion resonances at masses of around 500 MeV \cite{Pena:1996tf}. 

\vspace{-0.05in} 

It is possible, of course, to construct phenomenological $NN$ potentials that, with a sufficently large number of parameters, give an accurate description of the $NN$ scattering data. However, OBE potentials have several important advantages.  First, they  provide a physical mechanism for the interaction between nucleons. This implies that the parameters in these models have a physical meaning, and that, at least in principle, they can be related to, or even be determined through other physical processes. Second, it is possible to construct consistent electroweak currents for systems interacting through OBE, since the underlying microscopic processes are known \cite{Gross:1987bu}.  With phenomenological potentials this construction is less straightforward  because there is no implied microscopic description of the flow of electroweak charges through a nuclear system.  Third, when OBE is used in a covariant  formalism without time ordering, effective three- and many-body forces are automatically generated from the off-shell couplings of purely two-body OBE \cite{Sta97,Sta97b}.   With phenomenological potentials three-body forces must be independently constructed. 
Finally, OBE models are relatively simple, and depend only on a moderate number of parameters. A quantitatively accurate OBE model represents a very economical description of the $NN$ interaction.

\vspace{-0.03in} 

OBE models also have their limitations. Since they are not fundamental interactions, their validity does not extend to very short distances where QCD should provide the correct description. In potential models, this unknown short-distance part of the interaction is usually parameterized phenomenologically through vertex form factors with adjustable parameters. These form factors also serve to regularize otherwise divergent loop integrals that appear when the kernel is iterated.  But parameters that describe the unknown short distance physics cannot be avoided;  even more fundamental  potential models derived from ChPT require subtraction constants to renormalize and absorb infinities arising from the unknown short range physics.  At fourth order, a potential based on ChPT will have at least 24 unknown subtraction constants (parameters) \cite{Entem:2001qp}.

\vspace{-0.03in} 

After early phase shift analyses by the VPI  group \cite{MacGregor:1968ng}, both the VPI \cite{SAID} and Nijmegen \cite{Stoks:1993tb}  groups obtained optimal values of $\chi^2/N_\mathrm{data} \approx 1$ after eliminating data sets from their analyses, based on statistical arguments about their incompatibility with other data sets \cite{Ber88}. The Nijmegen group also updated their OBE potential (Nijm78) to the new phase shift analysis, but they were unable to get the $\chi^2/N_\mathrm{data}$ of this 15 parameter model (now called Nijm93) below 1.87 \cite{Sto94}. In order to construct very accurate $NN$ potentials they abandoned a pure OBE structure and made several boson parameters dependent on individual partial-waves.  Similarly, the (almost) pure OBE potentials of the Bonn family, such as Bonn A, B, and C, were superseded by the realistic CD-Bonn, which also incorporates partial-wave dependent boson parameters \cite{Mac01}. The Argonne group also motivated their construction of  largely phenomenological potentials like AV18 by the apparent failure of the OBE mechanism (apart from the pion-exchange tail) to allow a perfect fit to the data \cite{Wir95}. 

The main objective of this letter is to show that within the Covariant Spectator Theory (CST) it is, in fact, possible to derive realistic OBE potentials, and that these require  comparatively few parameters. This somewhat surprising finding contradicts the earlier conclusion and common belief that the OBE mechanism is missing some important feature of the $NN$ interaction. Accurate OBE models  may provide a useful intermediate step between fundamental physics and experiment.

In CST \cite{Gross:1969rv,
Gross:1982ny}, the scattering amplitude $M$ is the solution of a covariant integral equation derived from field theory (sometimes referred to as the ``Gross equation'').  In common with many other equations, it has the form
\bea
M=V-VGM
\eea
where $V$ is the irreducible kernel (playing the role of a potential) and $G$ is the intermediate state propagator.  As with the Bethe-Salpeter (BS) equation \cite{Salpeter:1951sz}, if the kernel is exact and nucleon self energies are included in the propagator, iteration of  the CST equation generates the full Feynman series.  In cases where this series does not converge (nearly always!) the equation solves the problem nonperturbatively.  With the BS equation the four-momenta of all $A$ intermediate particles are subject only to the conservation of total four-momentum $P=\sum_{i=1}^A p_i$, so the integration is over $4(A-1)$ variables.  In the CST equation, all but one of the intermediate particles are restricted to their positive-energy mass shell, constraining $A-1$ energies (they become functions of the three-momenta) and leaving only $3(A-1)$ internal variables, the {\it same number\/} of variables as in nonrelativistic theory. Since the on-shell constraints are covariant, the resulting equations remain manifestly covariant even though all intermediate loop integrations reduce to three dimensions, which greatly simplifies their numerical solution and physical interpretation. This framework has been applied successfully to many problems, in particular also to the two- and three-nucleon system \cite{Gro92,Sta97,Sta97b}.

The specific form of the CST equation for the two-nucleon scattering amplitude $M$, with particle 1 on-shell in both the initial and final state, is  \cite{Gro92}
\begin{align}
&M_{12}(p, p'; P)=
\overline {V}_{12}(p, p'; P)\cr
&-\int \frac{d^3 k}{(2\pi)^3} \frac{m}{E_k} 
\overline {V}_{12}(p,k;P)G_2(k,P)
{M}_{12}(k,p';P)\, ,\qquad
\end{align}
where $P$ is the conserved total four-momentum, and $p, p'$, and $k$ are relative four-momenta related to the momenta of particles  1 and 2 by
$p_1=\sfrac12 P+p$, $p_2=\sfrac12 P-p$, and
\begin{align}
M_{12}(p, p'; P)&\equiv M_{\lambda\lambda',\beta\beta'}(p, p'; P)\cr
&=\bar u_\alpha({\bf p},\lambda){\cal M}_{\alpha\alpha';\beta\beta'}(p, p'; P)u_{\alpha'}({\bf p'},\lambda')\quad
\end{align}
is the matrix element of the Feynman scattering amplitude ${\cal M}$ between positive energy Dirac spinors of particle 1.   The propagator for the off-shell particle 2 is
\bea
G_2(k,P)\equiv G_{_{\beta\beta'}}(k_2)=\frac{\left(m+\slashed{k}_2\right)_{_{\beta\beta'}}}{m^2-k_2^2-i\epsilon}\,h^4(k_2^2)
\eea
with $k_2=P-k_1$, $k_1^2=m^2$, and $h$ the form factor of the off-shell nucleon (related to its self energy), normalized to unity when $k_2^2=m^2$.  In this paper we use
\bea
h(p^2)=\frac{(\Lambda_N^2-m^2)^2}{(\Lambda^2_N-m^2)^2+(m^2-p^2)^2}\, ,
\eea
where $\Lambda_N$ is an adjustable cutoff parameter.
The indices 1 and 2 refer collectively to the two helicity or Dirac indices of particle 1, either  $\{\lambda\lambda'\}$ or $\{\alpha\alpha'\}$, and particle 2, $\{\beta\beta'\}$.

The covariant kernel $\overline V$ is explicitly antisymmetrized.  In its Dirac form it is
%
%\begin{widetext}
\begin{align}
&{\overline V}_{\alpha\alpha';\beta\beta'}(p,k;P)\cr
&=\sfrac12
\left[ V_{\alpha\alpha';\beta\beta'}(p,k;P)+(-)^I V_{\beta\alpha';\alpha\beta'}(-p,k;P)
\right] \, ,
\end{align}
%
%\end{widetext}
%
where  the isospin indices have been suppressed, so that the factor of $(-)^I$ (with $I$=0 or 1 the isospin of the $NN$ state) insures that the remaining amplitude has the symmetry $(-)^I$ under particle interchange $\{p_1,\alpha\}\leftrightarrow\{p_2,\beta\}$ as required by the generalized Pauli principle.  This symmetry insures that identical results emerge if a different particle is chosen to be on-shell in either the initial or final state.

Next we assume that the kernel can be written as a sum of OBE contributions 
\bea
V^b_{12}(p,k;P)=\epsilon_b\delta\frac{\Lambda^b_1(p_1,k_1)\otimes \Lambda^b_2(p_2,k_2)}{m_b^2+|q^2|} f(\Lambda_b,q)
\label{OBE}
\eea
with $b=\{s, p, v, a\}$ denoting the boson type, $q=p_1-k_1=k_2-p_2=p-k$ the momentum transfer, $m_b$ the boson mass, $\epsilon_b$ a phase, and $\delta=1$ for isoscalar bosons and $\delta=\tau_1\cdot\tau_2=-1-2(-)^I$ for isovector bosons.  All boson form factors, $f$, have the simple form
\bea  
f(\Lambda_b,q)=\left[\frac{\Lambda_b^2}{\Lambda_b^2+|q^2|}\right]^4
\eea
with $\Lambda_b$ the boson form factor mass.  The use of the absolute value $|q^2|$ 
amounts to a covariant redefinition of the propagators and form factors in the region $q^2>0$.  It is a significant new theoretical improvement that removes all singularities and can be justified by a detailed study of the structure of the  exchange diagrams.    The axial vector bosons are treated as contact interactions, with the structure as in (\ref{OBE}) but with the propagator replaced by a constant, $m_a^2+|q^2|\to m^2$ with a nucleon mass scale.
The explicit forms of the numerator functions $\Lambda^b_1\otimes \Lambda^b_2$ can be inferred from Table \ref{tab:Ls}.
Note that $\lambda_p=0$ corresponds to pure pseudovector coupling, and that the definitions of the off-shell coupling parameters $\lambda$ or $\nu$ differ for each boson.

\begin{table}
\begin{minipage}{3.2in}
\caption{Mathematical forms of the $bNN$ vertex functions, with $\Theta(p)\equiv(m-\slashed{p})/2m$.
The vector propagator is $\Delta_{\mu\nu}=g_{\mu\nu}-q_\mu q_\nu/m_v^2$ with the boson momentum $q=p_1-k_1=k_2-p_2$.}
\label{tab:Ls}
\begin{tabular}{lccl}
$J^P (b)\quad$ & $\epsilon_b$& $\quad\Lambda_1\otimes\Lambda_2\quad$  & $\Lambda(p,k)$ or $\Lambda^\mu(p,k)$ \cr
\tableline
$0^+ (s)$ & $-$& $ \Lambda_1 \Lambda_2$ &
$g_s-\nu_s\left[\Theta(p)+\Theta(k)\right] $\cr
$0^- (p)$ & + &$ \Lambda_1 \Lambda_2$ &
$g_p\gamma^5$\cr
&&&$-g_p(1-\lambda_p)\left[\Theta(p)\gamma^5+\gamma^5\Theta(k)\right]$\cr
$1^- (v)$ & $+$ & $\Lambda_1^\mu \Lambda_2^\nu \Delta_{\mu\nu}$
& $g_v\left[\gamma^\mu +  \frac{\kappa_v}{2M}i\sigma^{\mu\nu}(p-k)_\nu\right]$\cr
&&&$+g_v\nu_v \left[\Theta(p)\gamma^\mu  +  \gamma^\mu \Theta(k) \right]$\cr
$1^+ (a)$ & $+$  & $\Lambda_1^\mu \Lambda_2^\nu g_{\mu\nu}$ & $g_a\gamma^5\gamma^\nu$
\end{tabular}
\end{minipage}
\vspace{-0.2in}
\end{table}

In the most general case the kernel is the sum of the exchange of pairs of pseudoscalar, scalar, vector, and axial vector bosons, with one isoscalar and one isovector meson in each pair.   If the external particles are all on-shell, it can be shown that these 8 bosons give the {\it most general\/} spin-isospin structure possible (because the vector mesons have both Dirac and Pauli couplings, the required 10 invariants can be expanded in terms of only 8 boson exchanges), explaining why bosons with more complicated quantum numbers are not required.   By allowing boson masses (except the pion) to vary we let the data fix the best mass for each boson in each exchange channel.   Finally, we break charge symmetry by treating charged and neutral pions independently, and by adding a one-photon exchange interaction, simplified by assuming the neutron coupling is purely magnetic, $i\sigma^{\mu\nu}q_\nu$,  and  that all electromagnetic form factors have the dipole form. To solve the CST $NN$ equation numerically, it  was expanded in a basis of partial wave helicity states as described in \cite{Gro92}.

Previous models of the kernel, such as models IA, IB, IIA, and IIB of \cite{Gro92} and the updated, $\nu$-dependent versions such as W16 used in \cite{Sta97}, had been obtained by fitting the potential parameters to the Nijmegen or VPI phase shifts. In a second step the $\chi^2$ to the observables was determined. The models presented in this paper were fit directly to the data, using a minimization program that can constrain two of the low-energy parameters  (the deuteron binding energy, $E_d=-2.2246$ MeV, and the $^1S_0$ scattering length, $a_0=-23.749$ fm, chosen to fit the very precise cross sections at near zero lab energy).  This was a significant improvement, both because the best fit to the 1993 phase shifts did not guarantee a best fit to the 2006 data base, and because the low-energy  constraints stabilized the fits.   After the first fit was found, it would then be possible to vary the off-shell sigma coupling, $\nu_\sigma$, to give essentially a perfect fit to the triton binding energy.  However, the binding energies we report here were obtained from the best fit {\it without any adjustment\/}, confirming the results reported in Fig.~1 of Ref.~\cite{Sta97}.

\begin{table}
\begin{minipage}{3.5in}
\caption{Values of the 27 parameters  for WJC-1 with 7 bosons and 2 axial vector contact interactions.  All masses and energies are in MeV; other couplings are dimensionless;  $G_b=g_b^2/(4\pi)$.  Parameters in {\bf bold} were varied during the fit; those labeled with an $^*$ were  constrained to equal the one above.  The deuteron $D/S$ ratio is $\eta_D$, and the triton binding energy is $E_t$.  Experimental  values are in parentheses. }
\label{tab:par1}
\begin{tabular}{lcccrcc}
$b$ & I & $\quad G_b\quad$ & $m_b$ & $\lambda_b$ or $\nu_b$  & $\quad\kappa_v\quad$ & $\Lambda_b$\cr
\tableline
$\pi^0$ & $\quad1\quad$ & {\bf 14.608}&134.9766 &{\bf 0.153}$\,\,$& --- &{\bf 4400}\cr
$\pi^\pm$ & $1$ & {\bf 13.703} &139.5702 & ${\bf -0.312}$$\,\,$& --- &4400$^*$\cr
$\eta$ & $0$ & {\bf 10.684} & {\bf 604} & ${\bf 0.622}$$\,\,$& --- &4400$^*$\cr
$\sigma_0$ & $0$ & {\bf 2.307} & {\bf 429} & ${\bf -6.500}$$\,\,$& --- &{\bf 1435}\cr
$\sigma_1$ & $1$ & {\bf 0.539} & {\bf 515} & ${\bf 0.987}$$\,\,$& --- &1435$^*$\cr
$\omega$ & $0$ & {\bf 3.456} & {\bf 657} & ${\bf 0.843}$$\,\,$& ${\bf 0.048}$ &{\bf 1376}\cr
$\rho$ & $1$ & {\bf 0.327} & {\bf 787} & ${\bf -1.263}$\,\,$$& ${\bf 6.536}$ &  1376$^*$ \cr
$h_1$ & $0$ & {\bf 0.0026} & --- & ---$\,\,\,\,\,\,$& ---$\,\,$ &1376$^*$\cr
$a_1$ & $1$ & ${\bf -0.436}$ & --- & ---$\,\,\,\,\,\,$& ---$\,\,$ &1376$^*$\cr
\tableline
 \multicolumn{7}{c}{$\Lambda_N={\bf 1656};\; \eta_D=0.0256(1)  \,(0.0256(4));\; E_t=-8.48\, (-8.48)$}\end{tabular}
\end{minipage}
\vspace{-0.2in}
\end{table}

\begin{table}
\begin{minipage}{3.5in}
\caption{Values of the 15 parameters for  WJC-2 with 7 bosons.  See the caption to Table \ref{tab:par1} for further explanation. }
\label{tab:par2}
\begin{tabular}{lcccrcc}
$b$ & I & $\quad G_b\quad$ & $m_b$ & $\lambda_b$ or $\nu_b$  & $\quad\kappa_v\quad$ & $\Lambda_b$\cr
\tableline
$\pi^0$ & $\quad1\quad$ & {\bf 14.038}&134.9766 & 0.0& --- &{\bf 3661}\cr
$\pi^\pm$ & $1$ &  14.038$^*$&139.5702 & $0.0$& --- &3661$^*$\cr
$\eta$ & $0$ & {\bf 4.386} & 547.51 & $0.0$& --- &3661$^*$\cr
$\sigma_0$ & $0$ & {\bf 4.486} & {\bf 478} & ${\bf -1.550}$& --- &3661$^*$\cr
$\sigma_1$ & $1$ & {\bf 0.477} & {\bf 454} & ${\bf 1.924}$& --- &3661$^*$\cr
$\omega$ & $0$ & {\bf 8.711} & 782.65 & $0.0$& $0.0$ &{\bf 1591}\cr
$\rho$ & $1$ & {\bf 0.626} & 775.50 & ${\bf -2.787}$& ${\bf 5.099}$ &  1591$^*$ \cr
\tableline
 \multicolumn{7}{c}{$\Lambda_N={\bf 1739};\; \eta_D=0.0257(1)  \,(0.0256(4));\; E_t=-8.50\, (-8.48)$}
\end{tabular}
\end{minipage}
\vspace{-0.2in}
\end{table}

\begin{table}
\begin{minipage}{3.5in}
\caption{Comparison of precision $np$ models and the 1993 Nijmegen phase shift analysis.  Our calculations are in bold face.}
\label{tab:1}
\begin {tabular}{lcc|ccc} \multicolumn{3}{c}{models}&\multicolumn{3}{c}{$\chi^2/N_\mathrm{data}$}\cr
\tableline
Reference & \#\footnote[1]{Number of parameters} & year\footnote[2]{Includes all data prior to this year. } & $\;\;$1993$\;\;$ & 2000& 2007 \cr
\tableline
PWA93\cite{Stoks:1993tb} &39\footnote[3]{For a fit to both $pp$ and $np$ data.} &1993 &  0.99 & --- & ---\cr
&&&{\bf 1.09}\footnote[4]{Our fitting procedure uses the effective range expansion.  The Nijmegen $^3S_1$ parameters were taken from Ref.\ \cite{deSwart:1995ui}, but  as no $^1S_0$ parameters are available we used those of WJC-1.} &{\bf 1.11}&{\bf 1.12}\cr
Nijm I\cite{Sto94}&41\footnotemark[3]&1993&1.03\footnotemark[3]&---&---\cr
AV18\cite{Wir95} & 40\footnotemark[3] & 1995 &1.06 & --- & --- \cr
CD-Bonn\cite{Mac01} &43\footnotemark[3] & 2000 & --- &1.02&--- \cr
WJC-1& 27$\;$ & 2007 &{\bf 1.03}& {\bf 1.05} & {\bf 1.06} \cr
WJC-2 & 15$\;$ & 2007 & {\bf 1.09}& {\bf 1.11} & {\bf 1.12}
\end{tabular}
\end{minipage}
\vspace{-0.2in}
\end{table}

The parameters obtained in the fits are shown in Tables \ref{tab:par1} and \ref{tab:par2}.  The $\chi^2/N_\mathrm{data}$  resulting from the fits are compared with results obtained from earlier fits in Table \ref{tab:1}.  The data base used in the fits is derived from the previous SAID and Nijmegen analyses with new data after 2000 added. The current data set includes a total of 3788 data, 3336 of which are prior to 2000 and  3010 prior to 1993.  For comparison, the PWA93 was fit to 2514, AV18 to 2526, and CD-Bonn to 3058 $np$ data.   We restored some data sets  previously discarded because their $\chi^2$ were no longer outside of statistically acceptable limits, and this increased the $\chi^2$ slightly.  Phase shifts and a full discussion of the data and theory will be published elsewhere.  

In both of our models the high momentum cutoff is provided by the {\it nucleon\/} form factor and not the meson form factors.  Hence the very hard pion form factors merely reflect the fact that the nucleon form factors are sufficient to model the short range physics in the pion exchange channel.  The off-shell scalar couplings are perhaps the most uncommon features of these models.  They are clearly essential for the accurate prediction of three-body binding energies \cite{Sta97}.  It is gratifying to see that the pseudoscalar components of the pion couplings (proportional to $\lambda_p$) remain close to zero, even when unconstrained, and that effective masses of all the bosons remain in the expected range of 400-800 MeV.  

Aside from this, the parameters of WJC-2 are quite close to values expected from older OBE models of nuclear forces.  A possible exception is  the pion coupling constant, somewhat larger than the $g^2/(4\pi)=13.567$ found by the Nijmegen group.  The high-precision model WJC-1 shows some novel features: (a) $g_{\pi^0}>g_{\pi^\pm}$, (b) large $g_\eta$, and (c) small $g_\omega$.   

Why do these OBE models work so well? We are reminded of the Dirac equation; it automatically includes the $p^4/(8m^3)$ energy correction that contributes to fine structure, the Darwin term (including the Thomas precession), the spin-orbit  interaction, and the anomalous gyromagnetic ratio.   Similarly, the CST automatically generates  relativistic structures hard to identify, and impossible to add to a nonrelativistic model without new parameters.  

We draw the following major conclusions from this work:  (1) The reproduction of the $np$ data by the WJC-1 kernel is essentially as accurate as any other $np$ phase shift analysis or any other model.  This surprising result is achieved with only 27 parameters, fewer than used by previous high precision fits to $np$ data.   It remains to be seen whether the results will be equally successful once the $pp$ data are included.  (2)  Model WJC-1 gives us a new phase shift analysis, updated for all data until 2006, which is useful even if one does not work within the CST.
(3) The larger number of parameters of WJC-1 is not necessary unless one wants {\it very\/} high precision; model WJC-2 with only 15 parameters is also excellent and comparable to previously published  high precision fits. 
(4) The OBE concept, at least in the context of the CST where it can be comparatively easily extended to the treatment of electromagnetic interactions \cite{Gross:1987bu} and systems with $A>2$, can be a very effective description of the nuclear force.

%\vspace{0.05in}

{\bf Acknowledgements}: This work is the conclusion of an effort extending over more than a decade, supported initially by the DOE  through grant  No.~DE-FG02-97ER41032, and recently supported by Jefferson Science Associates, LLC under U.S. DOE Contract No.~DE-AC05-06OR23177. A.\ S.\ was supported by FCT under grant No.~POCTI/ISFL/2/275. We also acknowledge prior work by R. Machleidt and J.W. Van Orden, who wrote some earlier versions of parts of the $NN$ code.  The data analysis used parts of the SAID code supplied to us by R.\ A.\ Arndt.  Helpful conversations with the the Nijmegen group (J. J. de Swart, M. C. M. Rentmeester, and R.G.E. Timmermans) and with R.\ Schiavilla are gratefully acknowledged.

\end{document}